\documentstyle[12pt,epsf]{article}
\begin{document}
\begin{titlepage}

\newcommand\letterhead {%
\parbox{3cm}{}%\psfig{figure=/u/walet/TeX/Styles/UPLOGO.ps}}
\hfill\parbox{10cm}{   { \Large \it UNIVERSITY of PENNSYLVANIA}\\[0.2cm]
\hspace*{\fill}\parbox{6.5cm}{ Department of Physics\\
David Rittenhouse Laboratory\\
Philadelphia PA 19104--6396}}\\[0.5cm]
\bf PREPRINT UPR--0408T-R}
\noindent\letterhead
\vspace{2 cm}
\begin{center}
{\Large \bf Extracting Nuclear Transparency from p-A Cross sections}\\
\end{center}
\begin{center}
{\bf Sherman Frankel, William Frati, and Niels R. Walet}
\end{center}
\begin{center}
\vfill
{\em Submitted to Nuclear Physics A}
\end{center}
\end{titlepage}
\title{Extracting Nuclear Transparency from p-A Cross sections}
\author{Sherman Frankel, William Frati, and Niels R. Walet\\
Physics Department, University of Pennsylvania, Philadelphia, PA 19104}
\date{\today}
\maketitle
\begin{abstract}
         We study nuclear structure effects on the transparency in high
transverse momentum $(p,2p)$ and $(e,e'p)$ reactions.
We show that in the DWIA-eikonal approximation, even when correlations are
included, one can get a factorized expression for the transparency.
This depends  only on the average nucleon density $\rho(r)$ and
a correlation function. We develop a technique to include
correlations in a Monte-Carlo Glauber type calculation.
We compare calculations of $T$ using the eikonal formalism and
 a continuous density, with a
Monte Carlo method based on discrete nucleons.
\end{abstract}
%PACs numbers: 12.38 Aw, 12.38 Qk, 13.75.Cs, 13.85.Dz,25.40.Ve

\section {Introduction }

         High $p_t$ $(p,2p)$ interactions in nuclei have gained renewed
interest following the suggestion \cite{Brodsky} that fluctuations in
nucleon size may show observable effects due to color screening. Of
particular interest is the nuclear transparency, $T$, which is defined
as the probability that a proton traversing a nucleus makes one and only
one collision with a nucleon and that the emerging nucleons make no
further interactions. It can apply to any interaction, but in this paper
we consider the case where the basic interaction is either a proton-proton
or an electron-proton
interaction. The transparency is extracted from the experimentally
obtained quasielastic $(p,2p)$ events using the factorization assumption
\cite{Carroll,Heppelman}:
\begin{equation}
\frac{d}{dt} \sigma_{p-A} / Z = T \int
S(\vec{k},\epsilon,\sigma_{tot} ) \frac{d}{dt} \sigma_{\rm p-p}(s,t) d^3 k
d\epsilon
\label{eq:cross1}
\end{equation}
This is the expression found in Eq.~(2) of
Ref.~\cite{Heppelman}.
         In this factorized form $\sigma_{\rm p-p}$ is
the $(p,2p)$ cross section evaluated at the appropriate Mandelstam
variables, $s$ and $t$. The variable $s (k,\epsilon)$ is the on-shell
center-off-mass energy which is a function of the missing energy,
$\epsilon$, and missing momentum, -$\vec k$, measured from the kinematic
reconstruction of the $(p,2p)$ events, and $t$ is the measured 4-momentum
transfer. Finally $S(\vec k,\epsilon, \sigma_{tot})$ is a spectral
function, which depends on the proton-nucleon interaction cross
section, $\sigma_{tot}$, and on the spatial distribution of nucleons in
the nucleus. It is normalized to unity.

         Using Eq.~(\ref{eq:cross1}) $T$ can be extracted from the
experimental data by weighting each observed event with the reciprocal
of $\frac{d}{dt} \sigma_{\rm p-p}(s,t) $, to remove the dependence of $T$ on
$\vec k$ and $\epsilon$. For an experiment designed to accept events of any
momentum $\vec k$ and missing energy $\epsilon$
there is no need to know $S(\vec k,\epsilon, \sigma_{tot})$, since
$\int S(\vec k,\epsilon, \sigma_{tot}) d^3 k\,d\epsilon=1$.
 In practice, at incident momenta of 6-10 GeV which are much larger
than the average $k$, corrections for apparatus acceptance as a function
of $\vec k$ are not difficult to obtain, especially since $S$ falls rapidly
with
$k$.

         If there were no initial state or final state interactions of
the protons traversing the nucleus, $T$ would be unity. Thus $T$
measures these initial and final state interactions which are sensitive
to the total proton-nucleon cross section, $\sigma_{tot}$, which in turn
will depend on the magnitude of the color screening. One of the
important ideas in color transparency is that $T$ will increase with the
momentum transfer, $t$, observed in the $(p,2p)$ reaction. Thus we examine
how to determine $T(t)$ in a way that will depend least on theoretical
calculations of its magnitude. This is also of value since it bypasses
the strong sensitivity of $T$ to the nucleon density near the nuclear
edge. (In general only a peripheral reaction will have a single
scattering, so that edge effects are enhanced.) We examine
the special case of the $t$ dependence of the nuclear cross section as
measured in a $(p,2p)$ reaction where the exiting protons are observed at
the angles $\theta_1$ and $\theta_2$.

         In this paper we show that a factorized expression for $T$,
Eq.~(\ref{eq:cross1}), can be derived from the Distorted Wave Impulse
Approximation. By using the closure approximation, as employed by Lee and
Miller \cite{Lee}, the spectral function  $S$ becomes independent of $\epsilon$
 (we denote this simplified function by $J$ in this work).
Even with this additional approximation, the expression for the
cross section is rather complicated, see Ref.~\cite{Lee}.
However, we show that the transparency depends only
 the nuclear proton radial distribution $\rho(r)$.
The density $\rho$ can
be obtained from elastic $e-A$ scattering without recourse to knowledge
of the individual single particle wave functions.
After using closure the transparency $T$ turns out to be
related to the classical probability that there are no initial or final state
scatters accompanying the $(p,2p)$ high $p_t$ interaction on a single nucleon.
We
shall see later how to take into account both the momentum and missing energy
taken off by the undetected nucleons measured in a fully reconstructed $(p,2p)$
experiment.

         We also show how the effect of nucleon correlations,
which affect the calculation of $T$,
can be included in a previously developed Monte-Carlo formalism for
Glauber scattering, and compare various model assumptions with the data.

We also discuss the transparency in the $(e,e'p)$ reaction,
for which both theoretical \cite{Benhar} and experimental studies
\cite{McKeown} have been performed. The formalism developed by
Benhar {\em et al.\/} closely parallels the DWIA eikonal formalism
used by Lee and Miller.

\section{Transparency and the Distorted Wave Impulse
Approximation}

         Let us first review the expected cross section on the basis of
the plane wave impulse approximation (PWIA). This cross section can be
obtained by using the closure approximation to sum over final states and
using on-shell values of the Mandelstam variables, $s$ and $t$, for the
quasi-free interaction \cite{Lee}. The result of this quantum-mechanical
approximation is to obtain a simple limit: The cross section, for an
incident proton of energy, $E_0$, on a nucleon with wave function $\psi$
becomes the probability, $P(k)$, of finding that nucleon in the nucleus
with momentum $k$, times the $(p,2p)$ cross section
at the on-shell values of $s(E_0, \vec{k}, \epsilon)$ and the momentum
transfer, $t$. While $t$ is directly obtainable from the measured
momenta, $\vec k$ and $\epsilon$ refer to the momentum and binding energies in
the {\it initial} state which are not directly measurable.

         For interaction with a single nucleon of independent particle
wave function $\psi$, the result is well known \cite{Lee}:
\begin{equation}
\frac {d\sigma_{p-A}}{dt} /Z = \int P(k)\frac
{d \sigma_{\rm p-p}[ s(k,\epsilon, t)]}{dt} d^3 k. \label{eq:1}
\end{equation}

         In particular $P(k) = | F(k)|^2$ where $F(k)$ is the Fourier
transform of the nucleon wave function, $\psi(r)$,

\begin{equation}
F(k)= \int e^{i\vec k \cdot \vec r} \psi(\vec r) d^3 r . \label{eq:2}
\end{equation}

     It is useful to recall that, in Eq.~(\ref{eq:2}), the wave function,
$\psi(\vec r)$, must be normalized to unity in order that $\int
P(k)d^3 k$ give the proper total probability, i.e., $\int P(k) d^3 k
= 1 $.

     We next examine the cross section in the distorted wave impulse
approximation (DWIA). In this approximation the plane waves for the
incoming $(i=0)$ and outgoing $(i= 1,2)$
particles are replaced by distorted
waves of the form $e^{i(\vec p_i\cdot \vec r)} D_i(\vec r, \rho(r),
\sigma_{tot}, \vec p_i)$, where $D_i$ is a factor describing the distortion of
the plane wave.
It depends on the total p-nucleon cross section
$\sigma_{tot}$, the nuclear density $\rho$  and the
momentum of the $i$th proton $\vec p_i$.

For the case of a single filled orbital we find that $F(k)$ in
Eq.~(\ref{eq:2}) is replaced by $F^\prime(\vec k)$,
\begin{equation}
F^\prime(\vec k) = \int e^{i\vec k \cdot \vec r} D_0 D_1
D_2\psi(\vec r) d^3 r . \label{eq:3}
\end{equation}
Note that since in general the $D$'s are not rotationally invariant the
momentum
distribution no longer depends only on the magnitude of $k$ but depends on both
the transverse momentum, $k_t$, and longitudinal momentum, $k_z$.

In order for $P^\prime(k) = | F^\prime(\vec k) |^2 $ to be a
proper {\it effective} momentum distribution which includes the initial
and final state interactions, $F^{\prime}(k)$ must be normalized so that
$\int P^\prime(k)d^3 k$ is unity. The normalization factor needed in
Eq.~(\ref{eq:3})
is defined as $\sqrt T $, anticipating the result displayed in
Eq.~(\ref{eq:6}), and is given by:
\begin{equation}
 T = \int \ D_0 D_1 D_2\psi(\vec r)^* D_0 D_1 D_2 \psi (\vec
r) d^3 r,
\label{eq:4}
\end{equation}
so that
\begin{equation}
P^\prime (k,\sigma) = \left| \frac {\int e^{i\vec k \cdot \vec r}
D_0 D_1 D_2 \psi (\vec r) d^3 r }{\sqrt T }\right|^2.
\label{eq:5}
\end{equation}
We then find, in analogy with (\ref{eq:1}),
\begin{equation}
         \frac {d\sigma_{p-A}}{dt}
/Z =
 T \int P^\prime (k) \frac {d\sigma_{\rm p-p}
   [ s(k,\epsilon, t)]}{dt}
d^3 k                        .
\label{eq:6}
\end{equation}

The p-p cross section can be obtained from p-p data where
 $s$ is
determined from $k$ and $\epsilon$, which in their turn are
 obtained by kinematically reconstructing
each event under the assumption that all the accepted events are pure
$(p,2p)$ events with no initial and final state rescatterings.

%This
%reconstruction requires measurement of the momenta of the two outgoing
%protons~\cite{recon}.

\begin{figure}
\epsfysize=7cm
\centerline{\epsffile{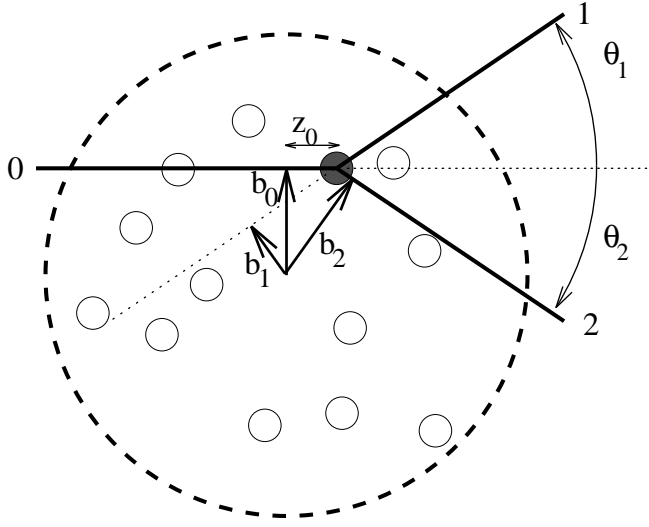}}
\caption{ Geometry of the scattering process.
The vectors $\vec b_i$ are the impact parameters
for each of the three paths (0 labels the incoming path, 1 and 2 label the
two outgoing paths) for a collision taking place at a nucleon at
 $\vec r$. $z_0$, and equivalently $z_1$ and $z_2$, are the
$z$ coordinates of $\vec r$ in the $\vec b, z$ coordinate systems.
$\theta_1$
 and $\theta_2$ are the angles of
the outgoing protons relative to the incoming proton direction.
\label{fig:defz}} %1
\end{figure}
In order to obtain the DWIA expression for $T$
we follow Lee and Miller and use the eikonal form of $D$ to evaluate $T$.
With the path lengths, $z$, defined in Fig.~\ref{fig:defz} ($z_0$ along the
direction of the incoming particle, $z_1$ and $z_2$ along the trajectories of
the outgoing particle) we have:
\begin{eqnarray}
D_0&=&\exp\left[{-\int_{-\infty}^{z_0} \frac {\sigma_{tot}}{2}
\rho(z,\vec{b}_0) dz} \right] ,
\nonumber\\
D_1&=&\exp\left[{-\int_{z_1}^ \infty \frac {\sigma_{tot}}{2}
\rho(z',\vec{b}_1)
dz'}\right]  ,
\nonumber\\
D_2&=&\exp\left[{-\int_{z_2}^\infty \frac {\sigma_{tot}}{2}
\rho(z'',\vec{b}_2)
dz''} \right] .
\label{eq:7}
\end{eqnarray}

Inserting these expressions in Eq.~(\ref{eq:4}) we obtain:
\begin{equation}
T = \int e^{-\int_{-\infty}^z   \sigma_{tot} \rho(\vec r) dz_0}
               e^{-\int_z^\infty        \sigma_{tot} \rho(\vec r) dz_1}
               e^{-\int_z^\infty        \sigma_{tot} \rho(\vec r) dz_2}
               \rho(\vec r) d^3 r = \int P_0 P_1 P_2 \rho(r)d^3 r.
\label{eq:8}
\end{equation}
The functions $P_i$, that are the squares of the functions $D_i$ in
(\ref{eq:7}), can be interpreted as the probability that there are
no interactions in the leg $i$ of the reaction.

In general we have several (partially occupied) orbits in a
shell-model nucleus. The result found here remains true, however.
Let us consider the example of two completely filled levels, $s$ and $p$,
with equal occupation numbers.
We define the ``normalized to unity'' effective momentum distribution
for state $i$ ($i= s, p$) by
\begin{equation}
         f_i(k) = \left| \int e^{i\vec k \cdot \vec r} D_0 D_1 D_2\psi_i
(\vec r) d^3 r \right|^2 /~T_i,
\label{eq:9}
\end{equation}
with the normalization factor determined from:
\begin{equation}
           T_i = \int[( D_0 D_1 D_2\psi_i(\vec r)]^*[D_0 D_1 D_2 \psi_i
(\vec r)]d^3 r .
\label{eq:10}
\end{equation}
We then have the useful
form:
\begin{equation}
          \sigma_{p-A}/Z = (T_s +T_p)
\int \left[ \frac { f_s T_s + f_p T_p } {T_s +
T_p}\right] \times  \sigma_{\rm p-p}(s,t) d^3 k  .
\label{eq:12}
\end{equation}
Note that $T_s + T_p$ is identical with $T$ since
\begin{equation}
           T_s + T_p = \int (D_0 D_1 D_2)^2 [\psi_s(r)^* \psi_s(r) +
\psi_p(r)^*\psi_p(r)] d^3 r = \int (D_0 D_1 D_2)^2 \rho(r) d^3 r.
\label{eq:11}
\end{equation}

This result can easily be extended to more independent particle states
and is the basis for our result that only $\rho(r)$ is needed to
calculate $T$. Thus Eq.~(\ref{eq:11}) is identical with
Eq.~(\ref{eq:8}).

     Note also the useful relation for the momentum distribution,
resulting from our normalization of $f$:
\begin{equation}
\int J(\vec k, \sigma) d^3 k \equiv \int \left[\frac {f_s T_s+f_p T_p}
{T_s + T_p}\right]
d^3 k = 1       . \label{eq:13}
\end{equation}

     The final form can thus be written:
\begin{equation}
          \frac{d}{dt} \sigma_{p-A} / Z = T \int J(\vec k, \sigma )
\frac{d}{dt} \sigma_{\rm p-p}(s,t) d^3 k   .
\label{eq:14}
\end{equation}
Thus, as in the PWIA, the DWIA also leads to a cross section that
can be factored into an effective normalized momentum distribution,
$J(\vec k)$, and a $k$ independent quantity, $T$, called the transparency.
This result relies on the use of a closure approximation, so that there
are no interference terms in Eq.~(\ref{eq:12}).
Let us repeat this equation (\ref{eq:8}) once more, since it is one of the
important results of this paper: $ T = \int P_0 P_1 P_2 \rho(r)d^3 r$.
     Thus we recognize that the normalization factor in
Eq.~(\ref{eq:4})   is in fact
just the transparency in Eq.~(\ref{eq:8})!
 This equation is exactly the expression for $T$
used by Farrar {\em et al.\/} \cite{Farrar} in  their
calculations.

Eq.~(\ref{eq:8})
can be understood as the probability that there is no nuclear
interaction of the incoming or outgoing protons along the paths $z_0,
z_1$ and $z_2$ determined by $\vec r$, $\theta_1$, and $\theta_2$.
However it does not include ``nuclear correlation''  effects that modify
$\rho(r)$ in Eq.~(\ref{eq:8})
in the neighborhood of the struck nucleon. These
corrections were calculated in Ref.~\cite{Frati} and
 are discussed and included by Lee and Miller~\cite{Lee}.
We have also calculated this probability using a Monte Carlo method
which automatically \cite{Frati} excludes the nucleon participating in
the high $p_t$ $(p,2p)$ interaction from the absorptive path.

Equations (\ref{eq:8}) and (\ref{eq:14})
allow $T$ to be determined directly from the proton
charge density, $\rho(r)$, which can be obtained from elastic e-nucleus
scattering rather than requiring a detailed knowledge of the ground-state
wave functions for all the nucleons of a complicated nucleus, which is the
procedure followed in Ref.~\cite{Lee}.

The correlations of the struck nucleon with its surroundings make
it less plausible to find another particle very close to this nucleon.
This can be most simply be taken into account by replacing the one-body
density $\rho$ in all three equations (\ref{eq:7}) by the probability
to find a particle at position $\vec r_{zb}$ along one of the three legs,
if there is one at the beginning (at position $r$),
\begin{equation}
\rho_2(\vec r,\vec r_{zb})/\rho(r) \equiv \rho(r_{zb}) C(\vec r,\vec r_{zb}).
\end{equation}
Here we have defined a correlation function $C$. This entity
is usually approximated by the result
of a nuclear matter calculation. As is argued in the appendix, however,
this correlation function should also take into account
the $(A-1)/A$ corrections arising from the fact that
one should not include the struck nucleon among the absorbing material.
Thus, for no correlations, we find $C=(A-1)/A$, which for
light nuclei gives a considerable correction to the transparency.

\section{The ``Glauber'' Calculation of $T$}
\begin{figure}
\epsfysize=7cm
\centerline{\epsffile{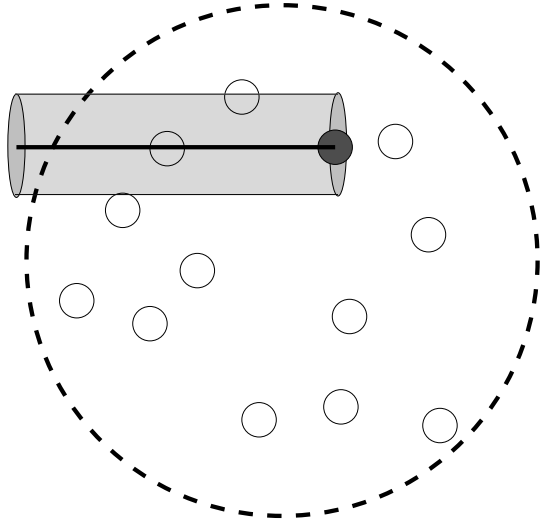}}
\caption{The geometrical cylinder where no particles are allowed
for a transparent reaction. For the calculation of
$a_n$ we draw a cylinder through the whole nucleus and
count the number of particles inside. Here we have illustrated
one of the legs in the $(p,2p)$ scattering, centered
on the particle where the hard collision is made.
We construct a similar
cylinder for each of the three legs.\label{fig:collision}}
\end{figure}

The Glauber model has proved to be quite successful in  understanding
p-A and heavy ion interactions at high energies ($>$ 5 GeV).
The basic quantities  such a calculation
are the so-called Glauber coefficients, $a_n$.
The coefficient $a_n$ gives the probability that a
proton makes $n$ collisions in traversing a nucleus, while
remaining on a straight path (this assumption is similar to the eikonal
approximation). One approximate way for obtaining the Glauber coefficients
is based on a Monte-Carlo generations of particles in a nucleus.
One  first generates the positions of
nucleons in a nucleus randomly, according to the nuclear density
$\rho(r)$.
One then follows the path of the incoming proton through the nucleus,
and counts the number of collisions. The collisions are treated geometrically,
and are assumed to take place whenever the
impact parameter for the incoming proton and any target nucleon
is less than  $b_\sigma$ obtained from the proton-nucleon
cross section, with $\sigma_{tot} = \pi b_\sigma^2 $.
The number of collisions is thus equal to the number of centers
of particles in a cylinder of radius $b_\sigma$ around the path
of the projectile,
 see Fig.\ \ref{fig:collision}.
Here we have taken into account the fact that
at very high energies
the path of the incoming particle through the nucleus
can be approximated by a straight line.
Such a probabilistic calculation of
the $a_n$ is known to be equivalent to the
quantum mechanical Glauber model in the case of
factorized ground state densities \cite{Bialas} and has been used over
the last decade to determine the $a_n$ which are used in models that
study multiplicities \cite{FF,Ott}, $E_t$ distributions \cite{Ak,Brody},
leading hadron distributions \cite{Fra1} and many other measured
quantities in p-A and heavy ion collisions at high energies.

         For our transparency calculations we need determine the probability
that if the projectile nucleon makes a large-angle $(p,2p)$ scatter,
no other soft collisions take place. We sum over all particles in the
nucleus, and assume that the hard-collision takes place at one
particle at a time. We then look at the incoming path, and also
 follow the paths of the two final state protons, and look
whether a geometric scattering takes place along each path.
We allow  the nucleons to scatter into non-zero angles as
determined by the $(p,2p)$ kinematics.
The calculation is similar but not identical with the eikonal-DWIA
calculation discussed  in the previous section. The expression
for $T$ resembles Eq.~(\ref{eq:8}) with the exponential factor $P_i$
replaced by the probability that there are no other
interactions except the hard $pp$ interaction along the appropriate
path.

\begin{figure}
\epsfysize=7cm
\centerline{\epsffile{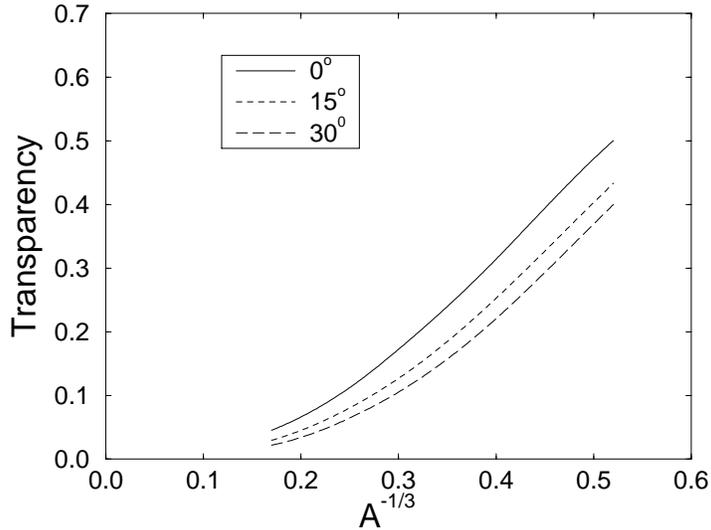}}
\caption{The transparency calculated as a function of $A$ for
the Glauber Monte Carlo model. The three curves represent
three different laboratory angles for the $pp$ scatter.\label{fig:ang}}
\end{figure}

In figure \ref{fig:ang} we show the dependence of the transparency
on the laboratory angle of the $pp$ scatter. The kinematics
are usually chosen so that the cm scattering angle is $90^\circ$,
we see that the transparency increases trivially as the energy
increases. We have used $\sigma_{\rm tot}=37~{\rm mb}$, which we
consider to be a reasonable value for the high-energy $pp$
cross-section in nuclei. This value will be used
in all calculations in this paper.

         One basic difference between this Glauber calculation and the
DWIA calculation is that the Monte Carlo treats the nucleus as composed
of confined nucleons of finite volume while the DWIA treats a continuous
matter distribution. We believe that at 10 GeV, where the wavelength of
the projectile is smaller than the nucleon diameter,
one cannot neglect the ``lumpiness'' of the nuclear medium.
In contrast to the eikonal calculation discussed in the
previous section, in the Glauber calculation the target nucleon
participating in the hard
$(p,2p)$ scatter is by definition not counted in the absorptive path.
This does
not mean that nuclear correlations do not have an effect on
the result of the Glauber calculation, but the effect can be
different. Let us investigate the generation of correlated nuclei
using our Glauber-Monte Carlo algorithm.

\subsection{Monte Carlo generation of particles}
As in the Eikonal-DWIA calculations we wish to use a nuclear matter
calculation as a guide to construct a correlated sampling of
a finite nucleus. There are several reasons why we cannot use a
correlation function directly as input to the calculation. The most
important reason is the fluid nature of the nuclear many-body system,
which induces many-particle correlations.

\subsubsection{Nuclear matter}
\begin{figure}
\epsfysize=7cm
\centerline{\epsffile{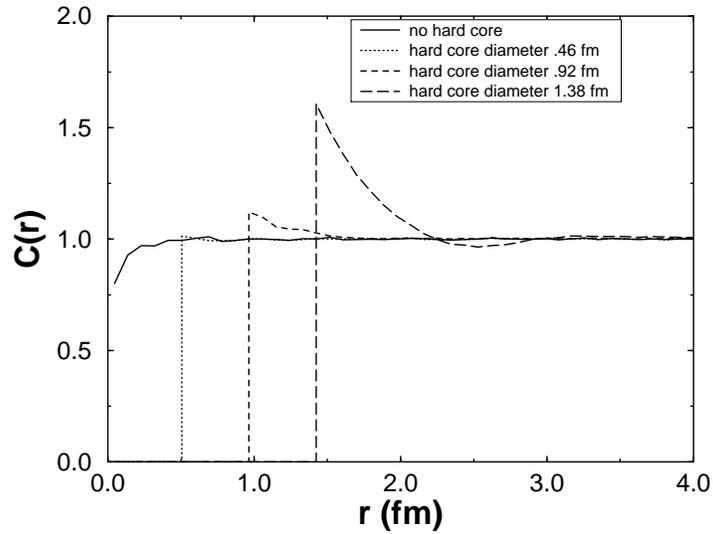}}
\caption{The effect of hard-core correlations in ``nuclear matter'', on the
correlation function $C$, for
various hard-core radii. The solid line is 0 hard core radius.
In sequence the other three lines have a hard-core diameter of .46,
.92 and 1.38 fm. \label{fig:gHardCore}}
\end{figure}

\begin{figure}
\epsfysize=7cm
\centerline{\epsffile{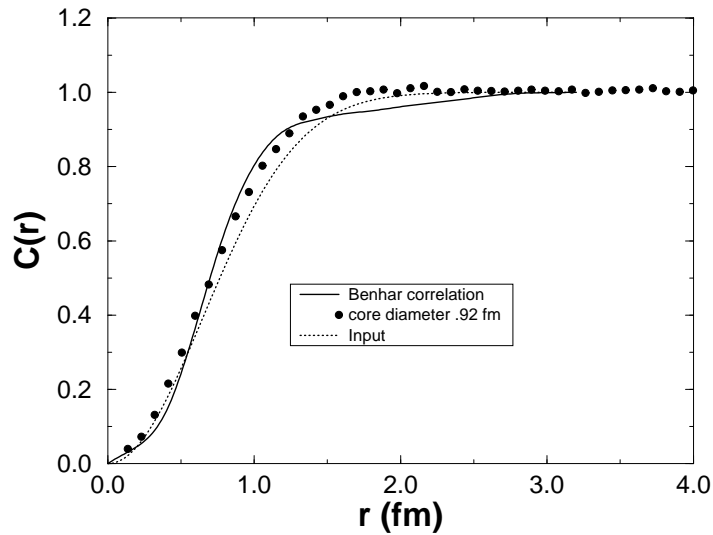}}
\caption{The effect of a Gaussian  correlation in ``nuclear matter'', on the
correlation function $C$.
The dashed line is the input Gaussian, the solid line the correlation function
used by Benhar {\em et al}. The black circles represent the outcome of
a Monte-Carlo calculation.
\label{fig:gGauss} }
\end{figure}

Let us first look at the case of nuclear matter.
If we consider particles with only two-body interaction, one expects
that the probability of finding the particles can be written as a
product of two-body correlation factors,
\begin{equation}
\rho(\vec r_1,\ldots) = \prod_i\rho \prod_{i<j} g(|\vec r_i -\vec r_j|)
\end{equation}
where $g$ is a two-body correlation factor. This function should approach
one  for large separations.  The point is that, for nuclear matter,
the correlation length -- the distance in which $g(r)$ approaches one --
is comparable to the internucleon spacing. This means that in general
we will be able to find triples of particles ($\vec r_i$, $\vec r_j$,
$\vec r_k$) in the Monte-Carlo sample
such that at least two of the three $g$'s ($g(\vec r_{i}-\vec{r_j})$,
$g(\vec r_{i}-\vec{r_k})$ and
$g(\vec r_{j}-\vec{r_k})$)
differ significantly from one. This leads to effective
(induced) three-body correlations.
If we now calculate the two-particle distribution,
the probability to find one particle at $\vec r_1$ and
another at $\vec r_2$,
\begin{equation}
\rho_2(\vec r_1, \vec r_2) = \rho^2 C(|\vec r_1 - \vec r_2|),
\end{equation}
we find that
the function correlation function $C$ is in general
different from $g$. (Note that in the low-density (gas) approximation $C=g$.)
This can be seen in Figs. \ref{fig:gHardCore} and \ref{fig:gGauss}
for two choices of the function $g$.
We have used
a Monte-Carlo algorithm where we generate N particles in a square box
with periodic boundary conditions.

In Fig.~\ref{fig:gHardCore} we show the effect of a hard-core interaction,
\begin{equation}
g(r) = \left\{\begin{array}{lr} 0 & r<R\\1 & r>R \end{array}\right. .
\end{equation}
As one can see the resulting curves do not have very much in common with
the results of more standard nuclear matter calculations
(see Fig.~\ref{fig:gGauss}).
The hard core reduces the available positions in configuration space
so that for the large core radii each particle is surrounded by
a number of particles that almost touch. This leads to the strong peak
in the correlation function for $r=R$.
In Fig.~\ref{fig:gGauss} we use a Gaussian $g$,
\begin{equation}
g(r) = 1-\exp(-(r/R)^2).
\label{eq:gGauss}
\end{equation}
This function results in a $C$ that -- for $R=.92~{\rm fm}$ -- is very
similar to the realistic correlation function  used by Benhar {\em et al.\/}
\cite{Benhar} in their calculation for the $(e,e'p)$ transparency.
This is what we use in our further calculations.

\subsubsection{Finite nuclei}
Whereas for infinite nuclear matter we can have an arbitrary,
but translationally invariant, correlation function,
the correlation function for finite nuclei must satisfy
a normalization condition.
This follows from the equation
\begin{equation}
\int d^3r_2 \rho^2(\vec r_1,\vec r_2) = (A-1) \rho(\vec r_2),
\end{equation}
which after the substitution
\begin{equation}
\rho^2(\vec r_1,\vec r_2) = \rho(\vec r_1)C(\vec r_1,\vec r_2\rho(\vec r_2)
\end{equation}
becomes
\begin{equation}
\int d^3r_2 C(\vec r_1,\vec r_2)\rho(\vec r_2)/(A-1) = 1.\label{eq:Norm2}
\end{equation}
We obviously also require that the one-body distribution coming
from the Monte-Carlo generation takes on the form $\rho(\vec r)$.

We wish to develop an approach to finite nuclei that is
close in spirit to the nuclear matter calculation discussed in the
previous subsection. Therefor we use an approach based on
two-body
correlation factors. We use a  $N$-particle distribution $\rho$ of the
form
\begin{equation}
\rho_{N}(\vec r_1,\ldots,\vec r_N) =
{\cal N}\prod_{i=1}^N \zeta(r_i) \prod_{i<j}^N g(|\vec r_i - \vec r_j|).
\end{equation}
${\cal N}$ is a normalization factor, that could be absorbed in the
effective single particle densities $\zeta$. The function $\zeta$ is chosen
such that $\rho(\vec r)$ takes a given form (usually Wood-Saxon),
whereas $g$ is taken to be identical to the nuclear matter result.
For a dilute gas one find that $\zeta =\rho$ and $g=C$. Unfortunately
a nucleus is more like a fluid than a gas, so that we have to solve
for $\zeta$.

In this  work we require that
for $g$ given by Eq.~(\ref{eq:gGauss}) we obtain a Wood-Saxon density,
\begin{equation}
\rho(r) \propto \frac{1}{1+\exp([r-R_0]/\sigma)},
\label{eq:rhoWS}
\end{equation}
with $\sigma = .545~\rm fm$ and $R_0 = 1.14 A^{1/3}~\rm fm$.
It is rather complicated to solve the many-body problem
for $\zeta$. We found it convenient to use a parametrized form
and look for a set of parameters that yields a result
that  closely resembles a Wood-Saxon.
A form that seems to be sufficient for this goal is
\begin{equation}
\zeta(r) \propto \frac{1}{\left(1+\exp([r-R'_0]/\sigma')\right)
\left(1+\alpha r\right)}.
\label{eq:zeta}
\end{equation}
The function $\zeta(r)$ is a parent distribution, from which
we chose particles in the Monte-Carlo calculation. If we include
the correlation factors $g$ (by a rejection technique) this leads to a
single particle distribution of the desired form $\rho(r)$.

\begin{figure}
\epsfysize=12cm
\centerline{\epsffile{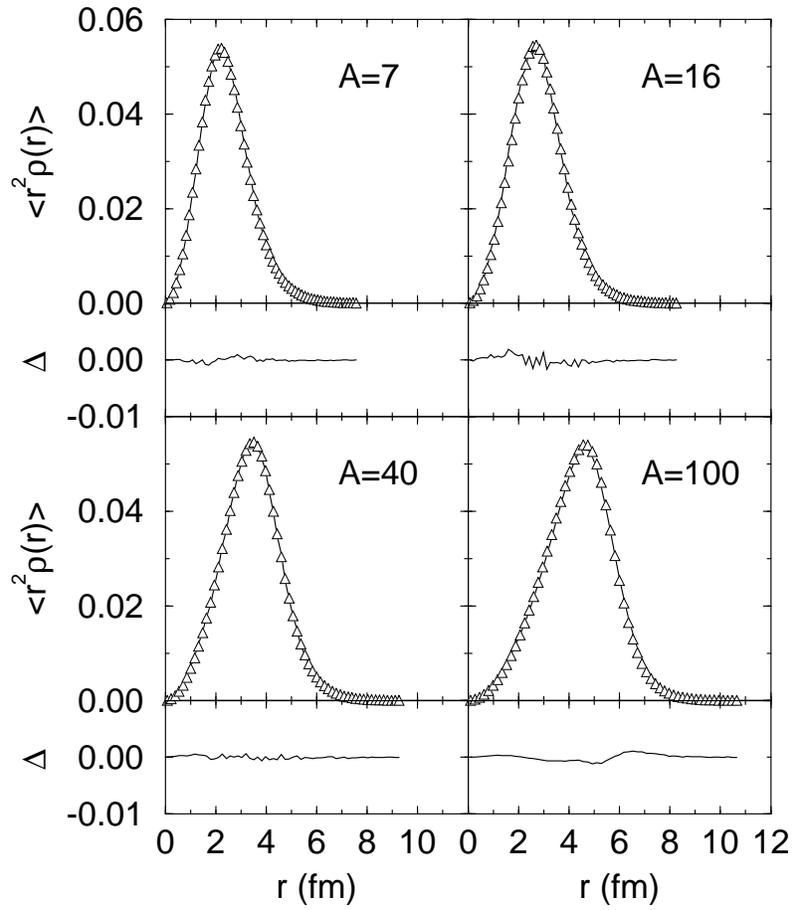}}
\caption{A comparison of the result of the density for the
correlated Monte-Carlo result to a Wood-Saxon, for four
different nuclei. For each pair of graphs the
upper one shows the value of the single-particle distribution
in a set of bins. The triangles are the Monte-Carlo results,
and the drawn line is the exact result for a Wood-Saxon distribution.
The lower graph shows the difference, $\Delta$, between the
two sets of results.
  For parameters see Table~\protect{\ref{tab:zeta}}.\label{fig:compn}}
\end{figure}

To illustrate the accuracy obtained by using (\ref{eq:zeta}) to arrive
at the Wood-Saxon distribution (\ref{eq:rhoWS}), we show a few of
the results of our calculations in Fig.~\ref{fig:compn}.
The relevant parameters for a few nuclei are given in Table \ref{tab:zeta}.
\begin{table}
\begin{center}
\caption{The parameters of $\zeta$, Eq.~\protect{\ref{eq:zeta}}
 for a few nuclei, as determined from a Monte-Carlo calculation.
\label{tab:zeta}}
\begin{tabular}{lllll}
$A$ & $R_0$ (fm) & $R_0'$ (fm) & $\sigma'$ (fm)& $\alpha$ (1/fm) \\
\hline
7  &  2.18 & 2.18 & .545 & 0.10 \\
12 &  2.61 & 2.55 & .545 & 0.13 \\
16 &  2.87 & 2.75 & .545 & 0.16 \\
40 &  3.90 & 3.70 & .545 & 0.19 \\
100 & 5.29 & 4.90 & .540 & 0.19 \\
208 & 6.75 & 6.17 & .530  & 0.19 \\
\end{tabular}
\end{center}
\end{table}

\section{Model Dependence of $T$}

\begin{figure}
\epsfysize=10cm
\centerline{\epsffile{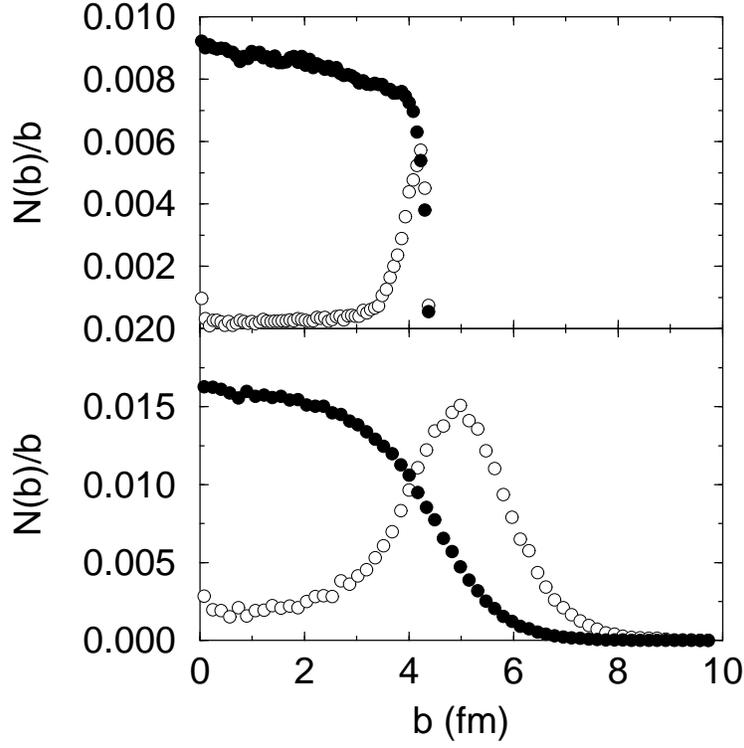}}
\caption{Impact parameter dependence of collisions in $^{64}$Cu.
Filled circles denote the number of all scatters
vs.\ the impact parameter, $b$.
Open circles give the number of transparent collisions multiplied by
ten. Both figures use  a Woods-Saxon density.
The upper figure has very small diffuseness, $\sigma=.005~{\rm fm}$,
the lower has a realistic diffuseness, $\sigma=.545~{\rm fm}$.
\label{fig:Nb_b}}
\end{figure}
         To elucidate the sensitivity of $T$ to the shape of the nuclear
surface we show in Fig.~\ref{fig:Nb_b} the dependence of $N(b)/b$,
the number of
collisions at impact parameter, $b$, versus $b$. Superimposed on the
same figure is the plot for the number of pure $(p,2p)$ events, i.e., those
without initial or final state scattering, $T(b)$ (multiplied by 10).
 These curves
illustrate how the tail of the $N(b)$ distribution determines both the
shape and position of $T(b)$. Nost of the transparent events take place
where the nuclear density is about $1/4$ of the maximum density.

\begin{figure}
\epsfysize=8cm
\centerline{\epsffile{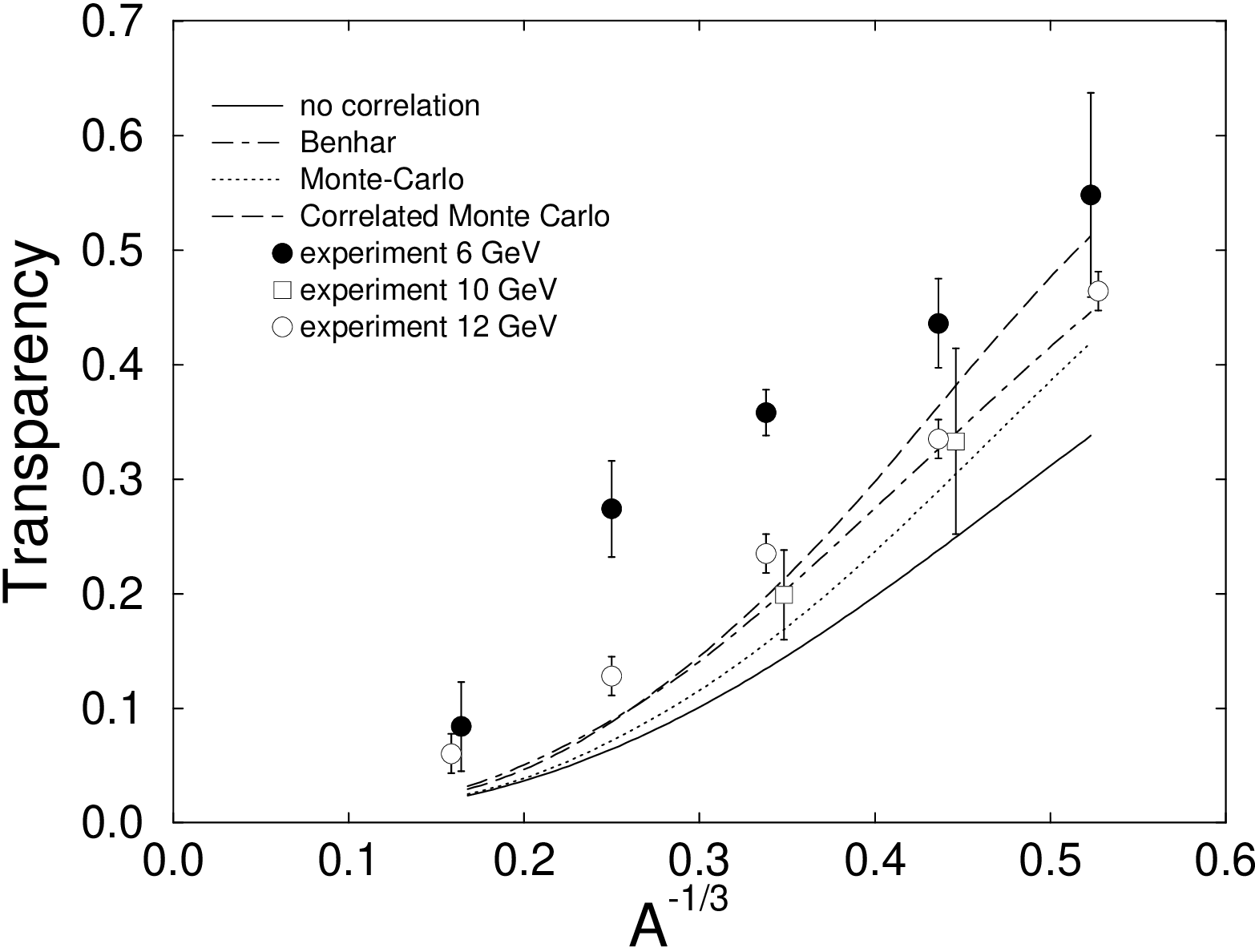}}
\caption{The transparency for the $(p,2p)$ reaction
calculated using various methods. The angles $\theta$ have been
chosen appropriate to a geometry corresponding to
$90^\circ$ scattering in the cm frame for $E_{lab}=10~{\rm GeV}$.
The 6, 10, and 12 Gev data are
from Ref.~\protect{\cite{Carroll}}.\label{fig:p2p}}
\end{figure}
In Fig.~\ref{fig:p2p} we compare the results of various calculational
schemes discussed in
this paper. We have calculated the transparency in $(p,2p)$ reactions
for the methods discussed in the previous sections,
all using a Wood-Saxon density of the form
\begin{equation}
\rho(r) \propto \frac{1}{1+\exp[(r-R_0)/\sigma]}
\end{equation}
where $R_0 = 1.14 A^{1/3}$ and $\sigma = .545~{\rm fm}$.
As stated before the total $pp$ cross section was
(in all cases) taken to be $\sigma_{rm tot}=37~{\rm mb}$.
We have calculated the transparency for the DWIA without correlations,
for the correlated DWIA-eikonal approximation, with the correlation
function used by Benhar {\em et al.\/} -- the one used by
Lee and Miller gives almost identical results --, the Glauber Monte
Carlo calculation without correlations and this last method with
correlations included.
The smallest transparency is found for the uncorrelated
DWIA calculation, comparable to the work of Farrar {\em et al\/}.
If we add correlations to the eikonal calculations, in the
form of a nuclear matter correlation function, the transparency
increases substantially. Similar results are found for the
Glauber calculation, where the results are
consistently higher than those from the eikonal calculation.
 The data of Carroll {\em et al.\/} \cite{Carroll} are superimposed
on the family of calculations in Fig \ref{fig:p2p}.

\begin{table}
\begin{center}
\caption{Transparencies a calculated for Fig.~\protect{\ref{fig:p2p}}.
The column labeled DWIA gives the Farrar-like results, DWIA~cor
gives the correlated eikonal-DWIA results with a Benhar type
correlation function. MC labels the uncorrelated Glauber Monte
Carlo calculations, and MC~cor gives the correlated result.
\label{tab:Transcomp}}
\begin{tabular}{lllll}
$A$ & $T_{\rm DWIA}$ & $T_{\rm DWIA~cor}$& $T_{\rm MC}$ & $T_{\rm MC~cor}$\\
\hline
7   & 0.342   & 0.446    & 0.419     & 0.512 \\
12  & 0.242   & 0.327    & 0.291     & 0.365 \\
16  & 0.197   & 0.270    & 0.233     & 0.292 \\
40  & 0.0965  & 0.134    & 0.108     & 0.136 \\
56  & 0.0732  & 0.101    & 0.0807    & 0.100 \\
100 & 0.0451  & 0.0611   & 0.0476    & 0.0575\\
208 & 0.0243  & 0.0319   & 0.0254    & 0.0298 \\
\hline
\end{tabular}
\end{center}
\end{table}

To ease the comparison for large $A$ we have also tabulated the
transparencies in Table \ref{tab:Transcomp}. From this table we see that
the correlated eikonal is, even for high $A$, 30\% larger than the
uncorrelated eikonal calculation. The effect of correlations
on the Glauber Monte-Carlo calculation depends somewhat stronger on $A$, and
is $22\%$ for $A=7$ and $17\%$ for $A=208$.

         We have found an error in the calculation of Ref.~\cite{Frati}.
 The effect of including correlations increases
the transparency by 60\% and not by 100\% as previously reported. As can be
seen from Table \ref{tab:Transcomp}  the Monte Carlo method gives somewhat
larger values of
$T$ than the eikonal approximation for a similar correlation function.

\begin{figure}
\epsfysize=8cm
\centerline{\epsffile{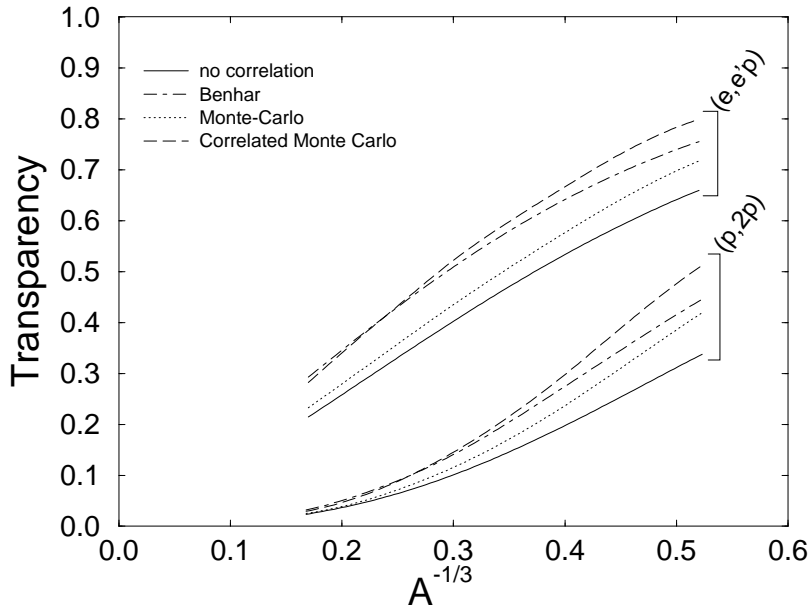}}
\caption{The transparency for the $(e,e'p)$ reaction
calculated using various methods. For ease of comparison
we have also included the $(p,2p)$ results.\label{fig:eep}}
\end{figure}
  Since our work was completed an important calculation of $T$ for the
(e,e'p) interaction has appeared in print \cite{Benhar}. Benhar et
al. have derived an expression for the transparency which carefully
incorporates the best present knowledge of the nuclear correlations.
The expression obtained in their work is very similar to the one
derived in Sec.~2. We have performed similar calculations,
corresponding to removing any of the particles in the nucleus
in a given direction. For the eikonal calculation
this leads to an expression
\begin{equation}
T = \int d^3r P_1(\vec r) \rho(r),
\end{equation}
and there is a similar simplification for the transparency in the
Glauber scheme.
  In fig.~\ref{fig:eep}
we compare the same  calculational schemes  as before.
For reference we also show the the $(p,2p)$ results. The difference in
curvature  as a function of $A$ can be explained as follows:
imagine the nucleus as a large sphere. If a proton comes in parallel
to the line connecting the poles, transparent events will only
take place near the equator (on the surface). The fraction of transparent
events is thus proportional to $R/R^3 \propto A^{-2/3}$. For the
$(e,e'p)$ reaction where the momentum transfer is parallel to
the same axis the whole southern hemisphere will contribute, leading to
a transparency proportional to $R^2/R^3 \propto A^{-1/3}$.

It is in our geometrical approach very natural to exclude the struck particle
from rescattering the projectile. In the original form of the eikonal model
this is not done. One should replace the one-body density $\rho$ in the
exponential by the ratio of the two-body and one-body density functions.
Lee and and Miller have modified their DWIA calculation to take into account
this effect approximately. This is done by multiplying $\rho(r)$ in
Eq.~(\ref{eq:7}) by a nuclear matter correlation function $C(z- z_i)$. The
correlation factor reduces the path length in the vicinity of the point where
the proton undergoing the $(p,2p)$ interaction is situated. They find that such
inclusion increases the value of $T$ by about a factor of 1.6 for
Carbon. Thus their result supports the remark \cite{Frati} that the
Farrar {\em et al.} calculation neglects this important effect. While
the increase in $T$ due to the correlations is similar in both our
calculations, the magnitudes may not be the same since the effective
correlation functions differ in the DWIA and probabilistic approaches.
Lee and Miller use a nuclear correlation function obtained from an
 earlier paper \cite{Miller}, which differs
from the nuclear correlation employed by ref.~\cite{Benhar}.

         Actually our results labeled ``Lee and Miller''
differ from those in their paper (\cite{Lee}) for another reason: The
definitions of $T$ are not identical. Lee and Miller define $T$ as
the ratio of the {\it theoretical} DWIA cross section to the {\it
theoretical} PWIA cross section which is not the same as the expression
used by the experimentalists (Eq.~(\ref{eq:14})) to extract $T$ from the
data. As we can see from Eq.~(\ref{eq:1}) and (\ref{eq:14}) their ratio,
$T_{LM}$, is given by
\begin{equation} T_{LM} = \frac {T \int J(\vec{k},
\sigma ) \frac{d \sigma_{p,p}(s,t)}{dt} d^3 k } {\int P(k)\frac {d
\sigma_{p,p} (s,t)}{dt} d^3 k}.
\label{eq:15}
\end{equation}
Since $J(\vec k,\sigma)$ is not equal to $P(k)$,
the Lee and Miller definition is not that used in the experimental
analysis.

\begin{table}
\begin{center}
\caption{Calculated Values of the transparency $T$ for
$^{12}$C in the $(p,2p)$ and $(e,e'p)$ reactions.
The kinematics used are appropriate to $90^\circ$ scattering
in the cm frame at a lab energy of $10~{\rm GeV}$.
Lee and Miller denotes the correlation function employed
in Ref.~\protect{\cite{Lee}}, Benhar that in Ref.~\protect{\cite{Benhar}}.
\label{tab:transpa}}
\begin{tabular}{lccll}
Method & $\rho (r)$ & Correlation& $T_{(p,2p)}$ & $T_{(e,e'p)}$\\
\hline
Eikonal  & Woods-Saxon & --- & 0.24&0.60\\
Eikonal  & Woods-Saxon & Lee and Miller  & 0.32&0.67\\
Eikonal  & Woods-Saxon & Benhar {\em et al.} & 0.33&0.68\\
Glauber  & Woods-Saxon & - & 0.29&0.63\\
Glauber  & Woods-Saxon & This work & 0.44&0.72\\
\hline
Eikonal  & Gaussian & - & 0.17&0.51\\
Eikonal  & Gaussian & Lee and Miller & 0.24&0.62 \\
Eikonal  & Gaussian & Benhar {\em et al.} & 0.25&0.63\\
Glauber  & Gaussian & - & 0.27&0.54\\
\hline
Eikonal  & $(e,e')$ & - & 0.18&0.53\\
Eikonal  & $(e,e')$ & Lee and Miller &  0.26&0.62\\
Eikonal  & $(e,e')$ & Benhar {\em et al.} & 0.27&0.63\\
Glauber  & $(e,e')$ & - & 0.28&0.55\\
 \hline
\end{tabular}
\end{center}
\end{table}

 The different predictions for Carbon are summarized in
Table \ref{tab:transpa}.
We have used a Wood-Saxon density as well as a Gaussian
density, obtained from using a completely
filled set of $s$ and $p$ orbitals with a length
parameter $b=0$, similar to the one employed by Lee and Miller.
We have also used an experimental
 charge-density obtained from  elastic electron scattering
 in these calculations \cite{NIKHEF}.
The results with this last density are close to the one obtained
with the harmonic-oscillator.
As can be seen for the $(p,2p)$ reaction the result of the
correlated eikonal and the Monte-Carlo Glauber calculation do
not differ very much. The effect of correlations on the Monte-Carlo
calculations leads to a result 33\% larger than the
correlated eikonal calculation. This result seems to be mostly
due to finite $A$ corrections, which are missing in the eikonal
calculation, and are important for this light nucleus.
It is important to note that the effect of correlations in
the Monte-Carlo calculations is to significantly enhance the transparency.
This shows that one should take into account two effects.
First one should not count the participants of the hard $pp$ scatter
among the absorbing volume. Secondly, the effect of correlations
on the local environment of the struck particle is ver important as
well.

         All of these effects, the sensitivity of $T$ to the shape of
the tail of $\rho(r)$, to the nuclear ``granularity'', or to the effect
of ``nuclear correlations'' in the eikonal model serve to emphasize that
the {\it magnitude} of $T$ may be a poor quantity to use for color
transparency studies.

\section{Extracting $T$ from the Data}

              We note from the defining equation, Eq.~(\ref{eq:14}),
that the cross section depends on three quantities,$S(k)$, $T$, and
$\sigma(s,t)$. If our goal is to search for the presence of color
screening, the $t$ dependence of $T$, and not its magnitude, is the
crucial quantity. Also, $S(k)$ has intrinsic interest, but not
necessarily to the extraction of $T(t)$.

         The experimental process for extracting $T$ from the data has
been given in Ref.~\cite{Carroll} and also in the thesis of Guang Yin
Fang \cite{Fang}. The experimentalists take the factorized form
(Eq.~(\ref{eq:14})) of the nuclear cross section as their basic
assumption which we see is the same as the DWIA result. To extract $T$
each observed event can be weighted with $W_{si}$ = $\frac{d}{dt}
\sigma(0,0) / \frac{d}{dt} \sigma(k,\epsilon) \equiv \frac{d}{dt}
\sigma_{p,p} / \frac{d}{dt} \sigma(k, \epsilon) $ to correct for the
cross section dependence on $k$ and $\epsilon$, and is also
divided by $Z$.
(Ralston and Pire \cite{Ralston} have pointed out the need to consider
variations of the $(p,2p)$ cross section from the simple $s^{-10}$ behavior.)

 If there were no acceptance biases, so that one could
use the fact that the integral over all $k$ is unity, $T$ would be
obtained directly.

         Since, strictly speaking, $T = T(\theta_1, \theta_2)$, and
since events with different $t$ might appear at different angles, one can
consider adding to future analyses the theoretical weighting factor,
$W_{1,2} = 1/ T(\theta_1 , \theta_2)$. Otherwise a $\theta$ dependence
on $t$ might appear as a spurious dependence of the transparency on $t$.
The rise in $T$ resulting in neglecting this effect can be read from
Fig.~\ref{fig:ang} for the special case of an equal angle configuration.

         Although we have emphasized that the magnitude of $T$ is model
dependent, the dependence of $T$ on the exiting angles would not be
expected to be. (The weight, $W_{1,2}$, is easily calculated from our
available Monte Carlo code.)
                   Thus we conclude that the data analysis can be
refined by using this angular weighting factor for more precise
determination of the average $T$ in the experiment.

         It might seem on a casual analysis that one could use the
transverse missing momentum spectrum observed from the data to predict
the longitudinal momentum distribution, since the transverse momentum
distribution is not sensitive to the strong $s$ dependence of the cross
section.
However whenever initial and final state interactions depend
on the projectile axis, the transverse and longitudinal momentum
distributions in $J(k,\sigma)$ are not identical. A nice internal check
of the data would be to unweight the $k_z$ distribution with the cross
section to see how much it differs from the transverse momentum
distribution. This is important if    the extraction of $T$ depends on a
knowledge of these distributions in order to correct for apparatus
acceptance as well as momentum cuts that might be applied to the data.

\section{Discussion}

         It is important to realize that the DWIA approximation results
in a factorized cross section which involves an effective momentum
distribution, $J(\vec k,\sigma)$.
The binding energy $\epsilon$ has disappeared from
this expression because of the sum over all final states in the closure
approximation. If the experiment were to detect all $(p,2p)$ events
independent of $k$ and $\epsilon$, this would not matter because that sum is
unity. But $S(\vec k ,\epsilon)$ is needed to carry out the Monte Carlo
calculations used to find the true apparatus acceptance. This is just
another way of saying that each event $i$ must be weighted by a $W_i(k,
\epsilon)$. Since protons of the same $k$ can be ejected from states
with different binding energies (and there is really a spectrum of
excitation energies for each initial state), it is clear that the
$S(\vec k, \sigma)$ appearing in the factorized formula should be
replaced by $S( \vec k, \epsilon, \sigma)$ so $\epsilon$ appears both in $S$
and in s. But knowledge of the complete spectral function will not be
needed if the apparatus is capable of accepting events of a wide enough
range of $\vec k$
 and $\epsilon$ so that the integral of $S$ will be close to unity.

         Finally, we point to an important experimental factor. While it
is important to determine $k$ and $\epsilon$ for each event to correct for the
strong $s$ dependence of the cross section, the key quantity determining
the number of pure $(p,2p)$ events, and hence $T$, is the efficiency of the
anticoincidence counters that are supposed to trigger on particles
produced by the inelastic interactions. If these detectors remove events
containing extra protons emitted because of the short range nucleon
nucleon correlations they will lower $T$. If they allow in an event
containing a pion, they will increase $T$, especially since such events
will be given a high weight when $\vec k$ is extracted from the
reconstruction. Thus complete knowledge of the anticoincidence
efficiency is crucial to extracting $T$. This requires generating the
soft collisions from the known event structures to find the efficiencies
at the different bombarding energies.

\section{\bf Conclusions}

         We have demonstrated, using the DWI approximation, that the
transparency depends directly on the nucleon spatial distribution $\rho
(\vec r)$ and the p,p total cross section. To calculate $T$ there is no
need to know the transparency for each nucleon wave function. We obtain
the classical result that the p-A cross section can be factorized into a
$k$ independent transparency, $T(\sigma)$, an effective ``normalized to
unity'' momentum distribution $S(k)$, and the p,p cross section
evaluated at the appropriate Mandelstam variables.

         Further we point out that the DWIA calculations of Lee and
Miller and our discrete nucleon calculation differ considerably in the
   magnitude of
     the ``nuclear correlations''.              In the latter we are
dealing with finite size nucleons distributed throughout the nuclear
volume rather than a smeared out uniform matter density, $\rho(r)$. The
nuclear correlation is incorporated simply, by not counting the struck
nucleon in the absorptive path. The smooth correlation function used by
Lee and Miller lowers the nucleon density near the struck nucleon in a
continuous way consistent with the DWIA. But calculation of their
$C(z-z^{\prime})$ shows that it makes for a smaller depression of $\rho$
than required by the existence of discrete nucleons. We suggest that one
should not apply the DWIA literally in computing $T$ at energies of ten
GeV where the proton wavelengths are smaller than the average
internucleon spacing and the confinement of nuclear matter into discrete
regions may better describe the physics.

\appendix
\section{Using a nuclear matter correlation function in finite nuclei.}

A nuclear matter correlation function cannot be added without
further ado to a finite nucleus calculation, since it does not have to
satisfy the condition (\ref{eq:Norm2}).
If we wish to remain as close in spirit to such a correlation function,
we need to require a simple renormalization.
The simplest (symmetric) way to renormalize the nuclear matter
$C^{\rm NM}(r)$ is to write
\begin{equation}
C(\vec{r}_1,\vec{r}_2) = \gamma(r_1) C^{\rm NM}(|\vec r_1 - \vec r_2|)
\gamma(r_2).
\end{equation}
This is symmetric under the interchange of $r_1$ and $r_2$, and we
can easily solve for $\gamma$ from the condition (\ref{eq:Norm2}).
This leads to the integral equation
\begin{equation}
\gamma(r_1) \int d^3r_2  C^{\rm NM}(|\vec r_1 - \vec r_2|)
\gamma(r_2)\rho(\vec r_2)/(A-1) = 1.\label{eq:inteq}
\end{equation}
We can easily see that even if there are no correlations, $C^{\rm NM}=1$,
$\gamma$ is not equal to 1, ($\gamma = \sqrt{\frac{A-1}{A}}$). This
shows the finite $A$ corrections missing in the original work
of Farrar {\em et al.\/} \cite{Farrar}.

The integral equation (\ref{eq:inteq}) can easily be solved numerically
on a grid in coordinate space. We have performed such calculations,
and find, as expected,
that the larger the nucleus, the smaller the effect. For
$^{12}$C $\gamma$ varies  by a few percent, whereas for
$^{208}$Pb the effect is much smaller.

\newpage

\end{document}